\documentclass[12pt,preprint]{emulateapj}
\newcommand\Rj{\mbox{$R_{\rm Jup}$}}
\newcommand\Mbol{\mbox{$M_{\rm bol}$}}
\newcommand\Msun{\mbox{$M_\sun$}}

\newcommand\maspix{mas~pix\mbox{$^{-1}$}}
\newcommand\Rhk{\mbox{$R^\prime_{HK}$}}

\begin{document}

\shortauthors{Metchev \& Hillenbrand}
\shorttitle{HD 203030B: an Unusually Cool L/T Young Dwarf}

\title{HD 203030B: an Unusually Cool Young Sub-Stellar Companion near the L/T Transition}
\author{Stanimir A.\ Metchev}
\affil{Department of Physics and Astronomy, University of California, Los Angeles, California 90095}
\email{metchev@astro.ucla.edu}
\and
\author{Lynne A.\ Hillenbrand}
\affil{Division of Physics, Mathematics \& Astronomy, California Institute of Technology, Pasadena, California 91125}
\email{lah@astro.caltech.edu}

\begin{abstract}

We present the discovery of a brown-dwarf companion 
to the 130--400~Myr-old G8~V star HD~203030.  
Separated by 11$\farcs$9 (487~AU in projection) 
from its host star, HD~203030B has an estimated mass of 
$0.023^{+0.008}_{-0.011}\Msun$.  The $K$-band spectral type of
L7.5$\pm$0.5 places HD~203030B near the critical L/T transition 
in brown dwarfs, which is characterized by the rapid disappearance 
of dust in sub-stellar photospheres.  From a comparative analysis 
with well-characterized field L/T transition dwarfs, 
we find that, despite its young age, HD~203030B has a bolometric luminosity 
similar to the $>$1~Gyr-old field dwarfs.  Adopting a radius from current 
models of sub-stellar evolution, we hence obtain that the effective 
temperature of HD~203030B is only $1206^{+74}_{-116}$~K,
markedly lower than the $\approx$1440~K effective temperatures 
of field L/T transition dwarfs.  The temperature discrepancy 
can be resolved if either: 
(1) the ages of field brown dwarfs have been over-estimated 
by a factor of $\approx$1.5, leading to under-estimated radii, or 
(2) the lower effective temperature of HD~203030B is related to 
its young age, implying that the effective temperature at the L/T 
transition is gravity-dependent.

\end{abstract}


\section{INTRODUCTION \label{sec_intro}}

After more than a decade of study, the fundamental parameters of brown dwarfs remain poorly constrained.  Unlike hydrogen-burning stars, whose effective temperatures and luminosities are approximately age-independent (at a fixed mass and composition) on the main sequence, brown dwarfs cool and dim continuously as they age.   Only one of the fundamental sub-stellar parameters, luminosity, can be estimated independently of the others, through the use of trigonometric parallax and empirically determined bolometric corrections.  Any attempts to resolve the remaining degeneracies among temperatures, ages, masses, and radii rely on the fortuitous discovery of brown dwarfs in association with stars or clusters of known age.  In such cases, brown-dwarf ages can be inferred by demonstrating the physical association of the brown dwarfs, usually as common proper motion secondaries or as members of stellar associations, with stars that have known ages.  Brown-dwarf masses can be measured from orbital astrometry and radial-velocity monitoring of close ($<$5--10~AU) binary systems, in which at least one component is sub-stellar.  Finally, brown-dwarf radii, and hence, effective temperatures and surface gravities, can be measured in sub-stellar eclipsing binaries.  The union of these three fortuitous cases, eclipsing brown-dwarf binaries associated with stars, offers the best opportunity for empirical determination of sub-stellar parameters.  However, such unusual systems are extremely rare---the first one, 2MASSJ~05352184--0546085, has only recently been reported \citep*{stassun_etal06}.  In all other cases involving brown dwarfs, the estimation of their properties relies to various degrees on the use of theoretical models of sub-stellar evolution \citep[e.g.,][]{burrows_etal97, burrows_etal01, chabrier_etal00, baraffe_etal03}.

Consequently, sub-stellar evolution models remain, for the most part, empirically unconfirmed.  Partial tests have been carried out for systems other than the ideal eclipsing double-line systems.  \citet{close_etal05} and \citet*{luhman_etal05} used astrometric measurements of the multiple system AB~Dor A/B/C to test the mass---age---luminosity relation at the stellar/sub-stellar boundary.  The two analyses offer examples of how sub-stellar evolution models can be tested without considering all degenerate sub-stellar parameters---in this case, excluding effective temperature.  We note, however, that the conclusions of the two teams differ \citep[see][for further discussion]{nielsen_etal05, luhman_potter06}, in part because the stellar parameters themselves (in particular, the stellar ages), although used as a reference in either study, may often not be known to the desired or believed level of accuracy.  In a separate example, \citet{mohanty_etal04b} and \citet*{mohanty_etal04} tackle the full set of degenerate sub-stellar parameters using high-resolution spectroscopic observations of brown dwarfs in the Upper Scorpius and Taurus associations.  Their studies rely on model spectra of brown-dwarf photospheres, constructed independently of sub-stellar evolutionary models, to resolve the mass---effective temperature---luminosity degeneracy.  Even though model-dependent, they allow a self-consistent comparison between sub-stellar cooling and photospheric models in the context of empirical data.

In the present paper we take a similar approach and discuss the degeneracies among sub-stellar age, luminosity, and effective temperature (i.e., excluding mass) in the context of models of sub-stellar evolution.  We address these fundamental properties with the help of a new brown dwarf that we discovered as a common proper motion companion to the 130--400~Myr-old main-sequence G8~V star HD~203030 (HIP~105232; \S\ref{sec_observations}).  The companion, HD~203030B, has a spectral type of L7.5, and thus lies very near the transition between L- and T-type sub-stellar photospheres, characterized by the settling of dust and the appearance of methane absorption in the near-IR spectra of brown dwarfs.  Because dust-settling occurs over a very narrow range of effective temperatures, 1500--1300~K (spectral types L6--T4), a fact inferred both theoretically \citep{ackerman_marley01, tsuji02} and semi-empirically \citep{golimowski_etal04, vrba_etal04}, the effective temperature of HD~203030B is expected to be constrained very well.  Given the known age and luminosity of HD~203030B, we can find its radius and thus offer a constraint on the theory (\S\ref{sec_properties}).  Our approach is not fully empirical, because it relies on the semi-empirical results of \citeauthor{golimowski_etal04} and \citeauthor{vrba_etal04}; both studies rely on evolution models for the sub-stellar age---mass---radius relations.  However, by adopting the same theoretical evolution models \citep{burrows_etal97, burrows_etal01} as in these two studies, our analysis provides a test of the self-consistency of the theoretical models (\S\ref{sec_lt_teff}).

\section{OBSERVATIONS: DETECTION AND FOLLOW-UP OF HD 203030B}
\label{sec_observations}

HD~203030B was discovered in the course of a direct imaging AO survey for sub-stellar companions conducted at the Palomar 5~m and Keck~II 10~m telescopes.  The observation strategy for the survey is described in \citet{metchev_hillenbrand04} and \citet{metchev06}.   Here we briefly describe aspects relevant to HD~203030.

\subsection{Imaging: Detection of HD 203030B}

HD~203030B was first detected at the Palomar 5~m telescope, using the facility AO system PALAO \citep{troy_etal00} and the Palomar High-Angular Resolution Observer \citep[PHARO; ][]{hayward_etal01} camera.  PALAO allows diffraction-limited (0$\farcs$09 at $K_S$-band) imaging in the near-IR with $K_S$-band Strehl ratios routinely above 50\% for natural guide stars brighter than $R=10$~mag.  PHARO employs a 1024$\times$1024~pix HgCdTe HAWAII detector, allowing a 25$\farcs$7$\times$25$\farcs$7 field of view (FOV) at a pixel scale of 25.09~\maspix.   We used a 0$\farcs$97-diameter opaque coronagraphic spot in PHARO to occult the primary in deep (60~sec) exposures.

HD~203030 was imaged at Palomar on three occasions: 28 August 2002, 16 July 2003, and 26 June 2004 (Table~\ref{tab_observations}).  During the first (discovery) epoch, we obtained a total of 24 individual 60-sec exposures of the primary with the coronagraph.  Subsequent epoch imaging was shorter (6--12~min), as dictated by the necessity to obtain only astrometric follow-up of the already discovered candidate companions.  In addition to the deep coronagraphic exposures, at each of the three epochs we obtained shallow (2--10~sec) non-coronagraphic $K_S$-band exposures of HD~203030 in a five-point dither pattern, to use as references for photometry and astrometry.  We also took five two-second $J$- and $H$-band unocculted exposures at first epoch, to measure the colors of the brightest candidate companions.  A 1\% neutral density (ND) filter was placed in the optical path in PHARO during the second and the third imaging epochs to prevent the primary from saturating in the unocculted $K_S$-band exposures.  The short $JHK_S$ exposures from the first epoch, however, were unattenuated and the primary was saturated.

A final (fourth) set of $JK_S$ imaging observations was obtained with the Keck II AO system \citep{wizinowich_etal00} on 12 July 2005.   On Keck we used the near-IR camera NIRC2 (K.~Matthews et al., in preparation) in its 20~\maspix\ scale and with a 1$\farcs$0-diameter partially transmissive coronagraph.   Both AO systems were employed in natural guide star mode, using HD~203030 ($V=8.4$~mag) itself as the guide star, which allowed $K_S$-band Strehl ratios around 50\% for all observations.

\begin{figure}
\plotone{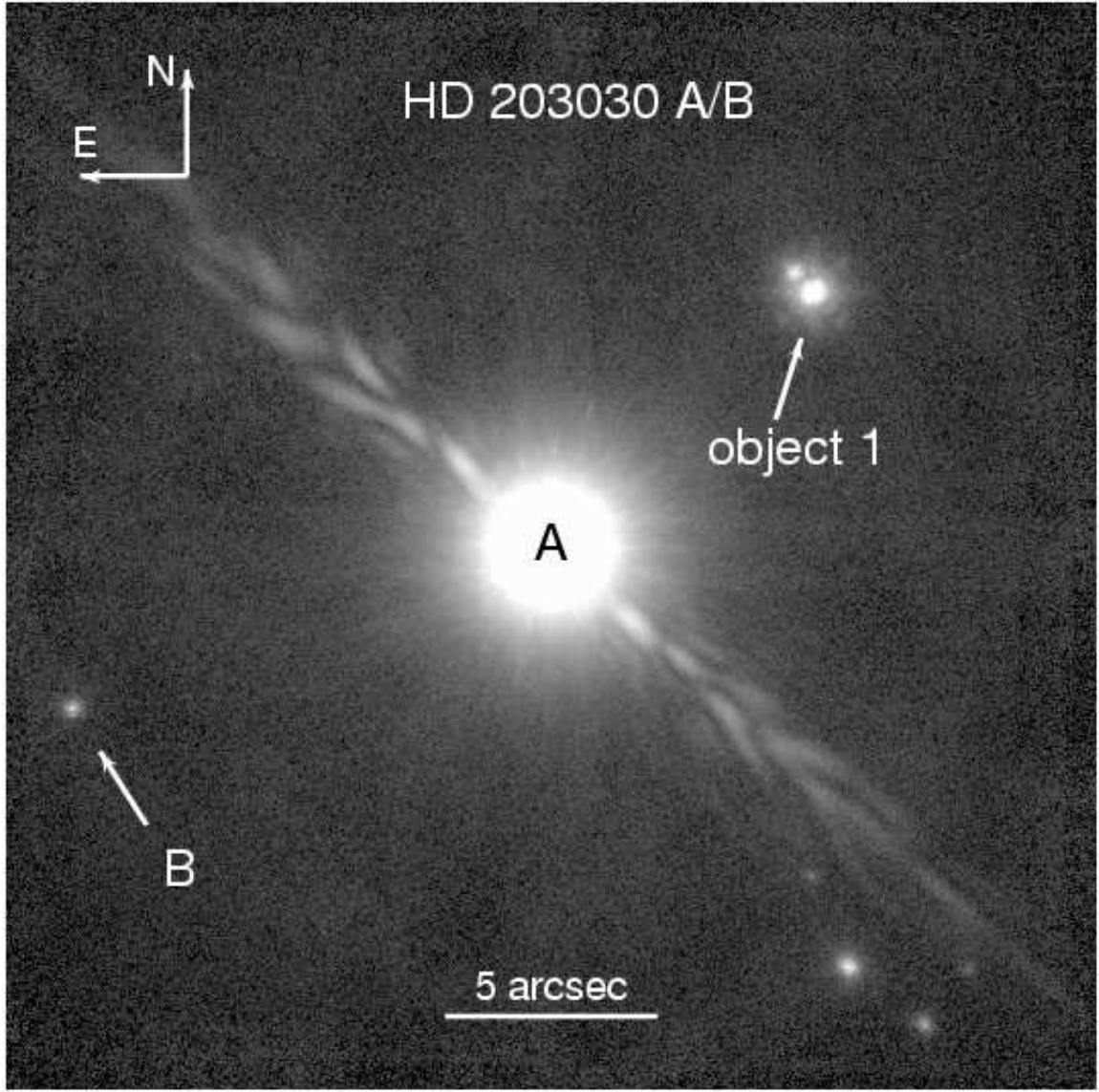}
\figcaption{The discovery image of HD~203030B, obtained at Palomar.  The total exposure time is 24~min at $K_S$ band.  The primary, HD~203030A is occulted by a $0\farcs$97 coronagraphic spot.  The secondary, HD~203030B is visible $11\farcs9$ (487~AU) away near the eastern edge of the image.  Six other background stars are also visible in the $25\farcs7\times25\farcs7$ field of view.  ``Object~1'' was used for bootstrapping photometric and astrometric measurements.  The wavy feature running diagonally across the image is an artifact due to a temporary oil streak on the secondary mirror of the telescope, present only in our 2002 data.
\label{fig_hd203030_image}}
\end{figure}

The reduction of the imaging data followed the standard steps of sky-subtraction, flat-fielding, and bad-pixel correction, as described in detail in \citet{metchev_hillenbrand04}.  The individual processed images were median-combined to produce a final high signal to noise image.   The final first-epoch image of HD~203030 is shown in Figure~\ref{fig_hd203030_image}.  A total of seven candidate companions are visible in the PHARO FOV, only one of which, HD~203030B (indicated with an arrow near the eastern edge of the image), was confirmed as a bona-fide proper-motion companion.  The brightest field object (``object~1'' in Fig.~\ref{fig_hd203030_image}) was a background star that was visible also in the short exposures and was used to bootstrap the relative photometry and astrometry of HD~203030B.

\subsubsection{Astrometry: Confirmation of Common Proper Motion \label{sec_astrometry}}

Measurements of the relative position of HD~203030B with respect to that of HD~203030A were obtained at all four imaging epochs.   Whenever possible, relative astrometry was obtained from images showing both the primary and the companion: the unocculted Palomar images from the first epoch, in which the primary was saturated, and the occulted Keck images from the last epoch, during which we used the partially transmissive NIRC2 coronagraph.  By fitting Gaussians to the PSFs of HD~203030A (or to its wings, if saturated) and of HD~203030B, we could measure each of their positions with 2--5~mas (0.04--0.10~pix) precision.  This was not possible for the images taken during the second and third Palomar epochs.  Because we used the PHARO ND filter for the short exposures at these epochs, the companion was not visible in the exposures in which the primary was unocculted.   And because the PHARO coronagraph is opaque, the primary was obscured in the deep coronagraphic exposures, in which the secondary was visible.  In the latter cases we estimated the position of the occulted primary by either: (1) bootstrapping its position in the coronagraphic images from the position of ``object~1'' (Fig.~\ref{fig_hd203030_image}) visible in the shallow exposures taken at the same epoch, or (2) inferring its position from the peak of the two-dimensional cross-correlation of the central $512\times512$~pix section of the coronagraphic image with a copy of itself inverted around its central pixel.  The latter technique takes advantage of the fact that the extended PSF halo of the primary is the dominant source of signal in the image, and that the halo is approximately radially symmetric on angular scales larger than the radius of the coronagraph.  Using either of these two approaches we constrained the position of the primary behind the coronagraph to within 13~mas ($\approx$0.5~pix).

At this precision level of astrometry, the dominant uncertainty is no longer measurement error, but systematic errors from the astrometric calibration.  In particular, accurate knowledge of the focal plane distortion on the detector is required.  The distortions of the PHARO 25~\maspix\ and NIRC2 20~\maspix\ cameras were mapped to high precision in \citet{metchev06} and \citet*{thompson_etal01}.  We applied the respective distortion corrections to the measured pixel coordinates, and calibrated the absolute pixel scales and orientations of the two detectors at each epoch from observations of WDS~18055+0230---a binary star with a well-known astrometric orbit from the Sixth Catalog of Orbits of Visual Binary Stars \citep{hartkopf_mason05}.

\begin{figure}
\plottwo{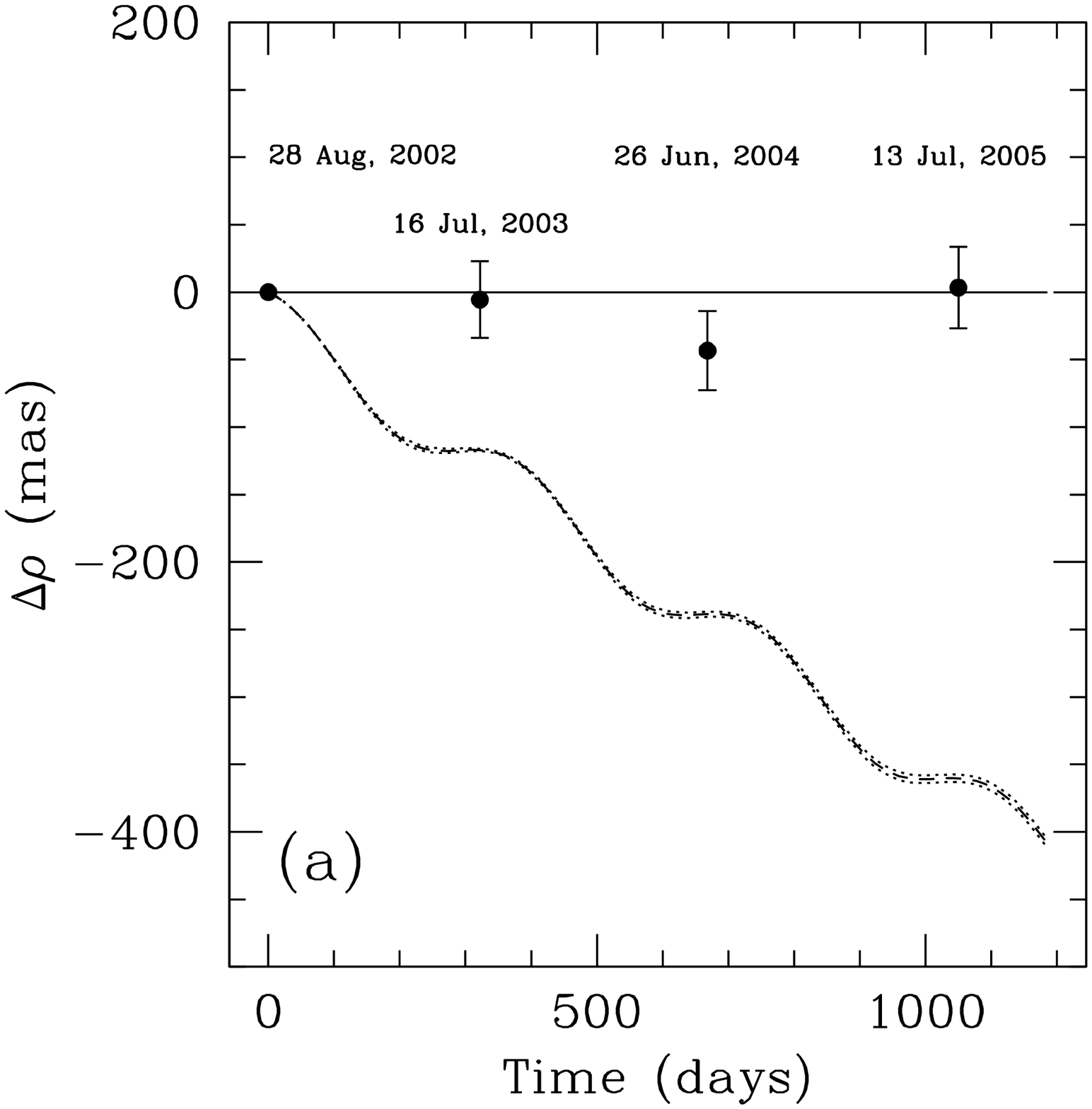}{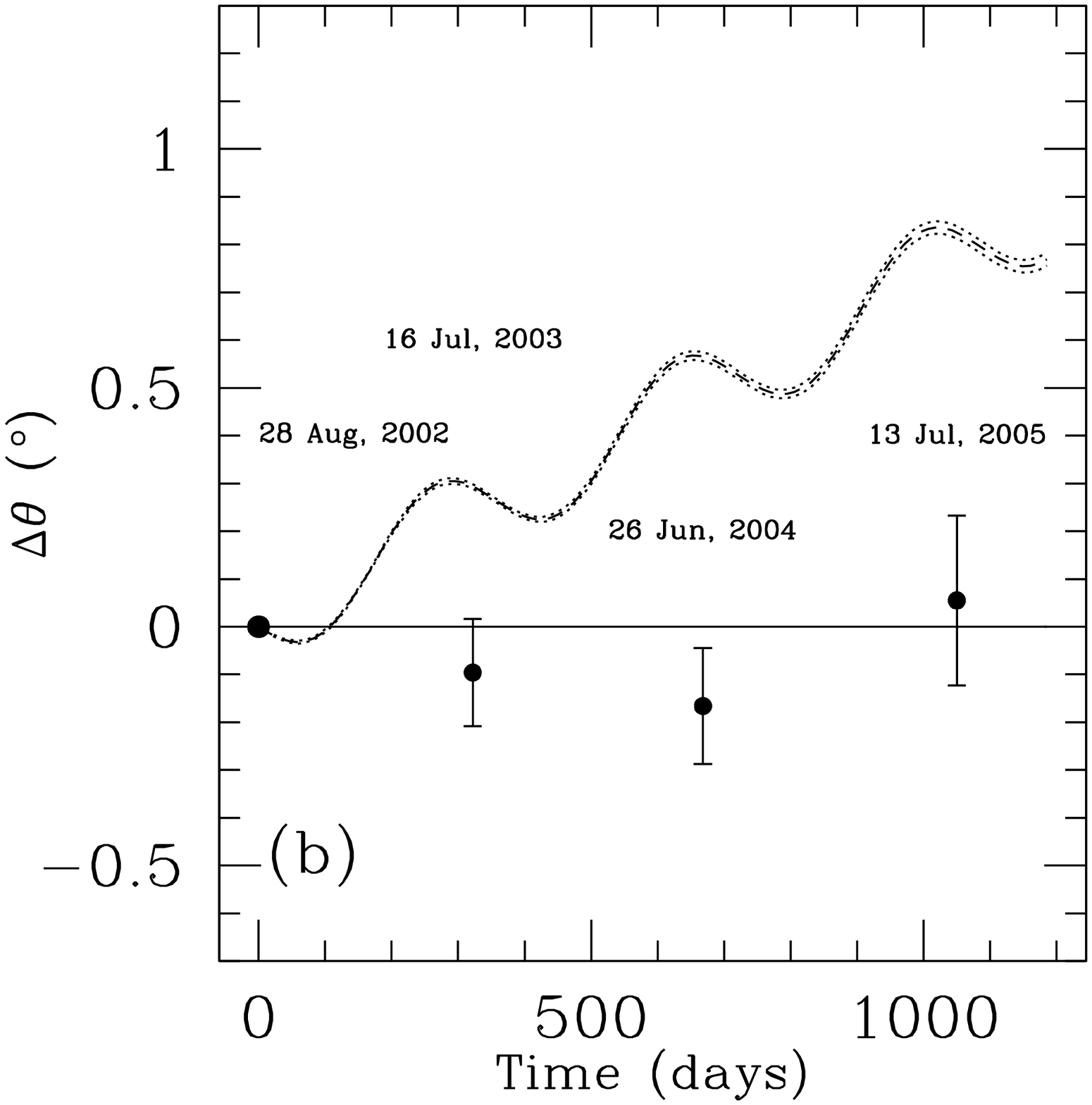}
\figcaption{Relative proper motion diagrams for HD~203030B, showing the change in radial separation (a) and position angle (b) with respect to the position of HD~203030A.  The astrometric measurements at the four observational epochs are denoted with dots.  The errors at the first epoch are not shown separately, but are absorbed into the errorbars of the measurements at the subsequent epochs.  The solid line in both panels traces null relative proper motion, i.e., common proper motion.  The dashed and dotted lines in the panels trace the expected relative motion of HD~203030B and its 1$\sigma$ error if the companion were an unrelated background star with negligible apparent proper motion.
\label{fig_hd203030_astrometry}}
\end{figure}

The four epochs of astrometry spanning nearly three years confirmed HD~203030B as a common proper motion companion of HD~203030A.  At first epoch, HD~203030B was separated by $\rho=11\farcs923\pm0\farcs021$ ($487.1\pm1.8$~AU) from the primary, at a position angle $\theta=108\fdg76\pm0\fdg12$.  The changes in the position angle ($\theta$) and separation ($\rho$) between the pair at subsequent epochs are shown with the data points in Fig.~\ref{fig_hd203030_astrometry}.  The astrometric errors of the first-epoch measurement (from 28 August, 2002) are added in quadrature to the errors of the measurements from the subsequent epochs.   The solid line in each of the panels in Figure~\ref{fig_hd203030_astrometry} represents the ``association'' line, synonymous with common proper motion.  The dashed lines in the panels trace the expected change in the relative position of HD~203030B with respect to HD~203030A, had the companion been a background star with a negligible apparent proper motion.  The dotted lines on either side of the dashed lines represent the $1\sigma$ uncertainties in the expected motion based on the {\sl Hipparcos} proper motion errors of the primary.  HD~203030B is inconsistent with being an unrelated ``stationary'' background star at the 13$\sigma$ level.  Because HD~203030 is not a member of a known open cluster, which could create the possibility of HD~203030B being just another kinematic member of the same cluster seen in projection, we assume that HD~203030A and B form a physical pair. 
 
\subsubsection{Photometry \label{sec_phot}}

Near-IR $H$ and $K_S$ magnitudes of HD~203030B were obtained from the individual unocculted 1.8~sec exposures taken at Palomar during the first epoch of observations (28 August 2002).  As the companion was not visible in the short $J$-band exposure, the $J$-band magnitude of HD~202030B was obtained from a deeper (50~sec) $J$-band coronagraphic image of HD~203030 taken with Keck AO on 14 July 2005, in which the companion was detected.

The primary was saturated in the short Palomar exposures and was occulted by the coronagraph in the Keck $J$-band exposures, thus preventing a direct measurement of the magnitude difference between HD~203030~A and B.  However, the night of 28 August 2002 was photometric, and we were able to calibrate the near-IR photometry with respect to another program star---HD~13531---unocculted $JHK_S$ images of which were obtained through the ND 1\% filter on the same night and at the same airmass.   HD~13531 is not known to be variable \citep[its {\sl Hipparcos} photometry is constant to within 0.015~mag;][]{perryman_etal97}, and is therefore an adequate photometric standard.  The near-IR transmissivity of the ND 1\% filter was established in \citet{metchev_hillenbrand04}.  The $H$- and $K_S$-band fluxes of HD~203030B and HD~13531 were measured in 0$\farcs$65-diameter apertures (encompassing 2 Airy rings at $K_S$).  Such wide apertures ($\approx$8 PALAO PSFs across) were chosen to mitigate as much as possible the anisoplanatic distortion (5--10\% in FWHM) of the PALAO PSF toward the edge of the 25$\farcs$7$\times$25$\farcs$7 PHARO field, and showed a high degree of self-consistency (r.m.s.$\leq$0.05~mag).  Experiments with smaller apertures, 0$\farcs$18--0$\farcs$36 in diameter, with sizes proportional to the (variable) FWHM of the PSF, reproduced our wide-aperture photometry to within 0.10~mag (though with a larger intrinsic scatter) without showing any systematic trends with aperture size.  Estimates of the local background were obtained from the mean of the pixel counts in 0$\farcs$25-wide annuli with inner radii between 0$\farcs$50 and 1$\farcs$25 (chosen to be at least 0$\farcs$25 larger than the radius at which the stellar radial profile blended into the background).  We used the 2MASS $H$ and $K_S$ magnitudes of HD~13531 to calibrate the transform from instrumental $H$ and $K_S$ magnitudes to apparent magnitudes.  

The $J$-band magnitude of HD~203030B was bootstrapped from the deeper NIRC2 images with respect to that of ``object~1'' (Fig.~\ref{fig_hd203030_image}).  We used the same 0$\farcs$65-diameter apertures, as for the $H$ and $K_S$ bands above.  At $J$-band the poorer AO performance (Strehl ratio $\lesssim$10\%) and the much more severe PSF elongation due to anisoplanetism limited the accuracy of our measurements.  All near-IR photometry of HD~203030B is listed in Table~\ref{tab_photometry}.

Finally, we note that HD~13531 itself was found to have a faint ($\Delta J=4.84$~mag; $\Delta K_S=4.20$~mag) stellar proper-motion companion at an angular separation of $0\farcs72$ \citep{metchev06}.   The presence of the secondary (HD~13531B) affects the 2MASS magnitudes of the primary by 0.01~mag at $J$ and by 0.02~mag at $H$ and $K_S$, which we take into account.  At its relatively large angular separation from HD~13531A (8--9 times the FWHM of the PALAO PSF), HD~13531B has a negligible effect on our photometric measurements.

\subsection{Spectroscopic Follow-up}

Spectroscopic observations of HD~203030B were obtained on 14 Jul 2005 with Keck AO after the association of the companion with the primary had been established.  With NIRC2 we used the 20~\maspix\ camera in conjunction with the 80~mas-wide slit and the {\sc lowres} grism for R$\sim$2000 $K$-band spectroscopy.  We obtained three 10~min $K$-band spectra at an airmass of 1.03--1.06 with dithers along the slit between exposures.  Conditions were photometric and the FWHM of the PSF as measured from the acquisition images was 50~mas.  Telluric and instrumental features in the spectra were identified and corrected from observations of a nearby B9.5V \citep{grenier_etal99} star, HD~201859, obtained immediately afterwards.  Images of internal flat field and arc lamps were taken at the beginning of the night to calibrate the detector response and the wavelength dispersion of the system.  

The individual frames were first flat-fielded and then pair-wise subtracted.  The spectra of HD~203030B were then extracted using the IRAF task {\sc apall}.  We adopted an aperture width of 14~pix ($0\farcs28$) centered on the spectrum of HD~203030B, which was traced by a high-order (6--10) polynomial.  We estimated the background in 40~pix ($0\farcs80$) wide bands centered 50--100~pix ($1\farcs0$--$2\farcs0$) away from the object.  The large widths of the slit and of the extraction aperture were chosen to minimize chromatic effects arising from the dependence of the FWHM of the AO PSF on wavelength \citep[e.g.,][]{goto_etal03}.  The success of this extraction approach was judged by the fact that three individual companion spectra extracted from wide apertures were consistent among each other, while variations in the AO PSF on time-scales of minutes are expected to induce chromatic differences among spectra extracted from narrower apertures.  Because HD~203030B is separated by almost 12$\arcsec$ from HD~203030A, stray light from the primary was not an issue in the extraction of the companion spectrum.  The spectra of the telluric standard were extracted in a similar fashion as those of HD~203030B, using the same aperture width, though with the background bands farther from the star to avoid contamination from the brighter PSF halo.

We calibrated the wavelength dispersion of the extracted spectra with the arc lamp spectra and, after interpolating over the intrinsic Br$\gamma$ absorption in the telluric standard, we divided the extracted spectra of HD~203030B by those of the standard.  We then median-combined the resulting three companion spectra and multiplied them by a 10500~K blackbody.  The final HD~203030B $K$-band spectrum was smoothed to match the instrumental resolution ($R\approx1300$, as measured from the FWHM of the arc lines) and is shown compared to spectra of known L6--T0 dwarfs in Figure~\ref{fig_hd203030b_spectrum}.

\subsubsection{Spectral Type of HD 203030B \label{sec_sptype}}

\begin{figure}
\plotone{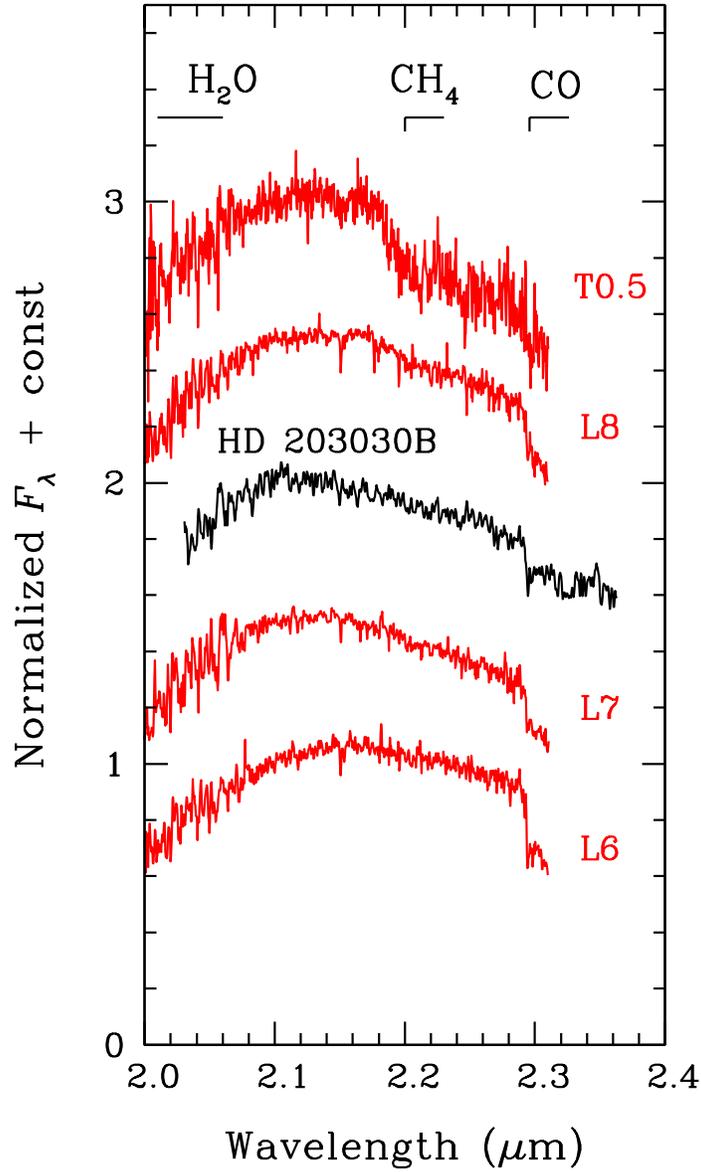}
\figcaption{A $K$-band $R\approx1300$ spectrum of HD~203030B obtained with Keck AO+NIRC2, compared to $R\approx2000$ spectra of L6--T0 dwarfs from \citet{mclean_etal03}.  The spectra have been normalized to unity at 2.1~$\micron$ and have been offset by a constant (0.5 in normalized flux) from each other for clarity.  HD~203030B shows H$_2$O and CO absorption, similar to other late L dwarfs, though does not show the characteristic CH$_4$ absorption of T dwarfs.  The comparison dwarfs from \citet{mclean_etal03} are 2MASS~01033203+1935361 (L6), DENIS-P~020529.0--115925AB (L7), GL~584C (L8), and SDSS~J083717.22--000018.3 (T0.5).
\label{fig_hd203030b_spectrum}}
\end{figure}

The $K$-band spectrum of HD~203030B (Fig.~\ref{fig_hd203030b_spectrum}) exhibits molecular absorption features due to H$_2$O in the blue end and CO in the red end, as is typical of late-M to T0 dwarfs.  No other molecular or atomic features are seen.  In particular, the spectrum does not show either \ion{Na}{1} doublet absorption at 2.21~$\micron$, which is seen in the spectra of $\leq$L2 dwarfs at similar ($R\sim2000$) resolution, or CH$_4$ absorption at 2.20~$\micron$, characteristic of $\geq$T0 dwarfs \citep{mclean_etal03}.  We thus infer that the spectral type of HD~203030B is intermediate between L2 and T0.  For a more precise spectroscopic classification of HD~203030B we rely on $K$-band spectroscopic indices from the NIRSPEC Brown Dwarf Spectroscopic Survey \citep[BDSS;][]{mclean_etal03}, designed to follow the L dwarf optical classification scheme of \citet{kirkpatrick_etal99}.

Following previous work on IR spectroscopic classification, \citet[][and references therein]{mclean_etal03} define three molecular indices in the $K$ band, measuring the onset of H$_2$O, CH$_4$, and CO absorptions at 2.05~$\micron$, 2.20~$\micron$, and 2.30~$\micron$, respectively.  The indices are defined as the ratios of the median flux values in 40~\AA-wide regions on either side of the molecular bandhead.  They are therefore applicable to data with somewhat lower than the original $R\approx2000$ resolution of the BDSS data, and are well-suited for our $R\approx1300$ NIRC2 spectrum.   We consider only the $K$-band CH$_4$  and CO indices of \citeauthor{mclean_etal03}, for which we measure values of $0.90\pm0.04$ and $0.85\pm0.03$, respectively, where the index means and their errors are obtained from measurements on the three independent HD~203030B spectra.  We do not use their $K$-band H$_2$O index because it requires measurements blue-ward of 2.0~$\micron$, where our spectrum of HD~203030B is cut off by the transmissivity of the NIRC2 {\sc lowres} grism.

Comparing to data from \citet{mclean_etal03} and \citet{cushing_etal05}, we find that the 1$\sigma$ spectral type ranges based on the CH$_4$B and CO indices are L7--T0 and M7--L8, respectively.   As already noted in \citeauthor{mclean_etal03} and \citeauthor{cushing_etal05}, the CO index is of little utility for spectral typing of ultra-cool dwarfs because it displays very little sensitivity to their effective temperatures.  Hence, we use only the result from the CH$_4$B index.  Having excluded the possibility that HD~203030B is a T dwarf, based on the lack of methane absorption at 2.20~$\micron$, we adopt a final spectral type of L7.5$\pm$0.5 for it on the L-dwarf classification scheme of \citet[][extending to L8 in the L dwarfs]{kirkpatrick_etal99}.  If, instead, we adopt the L-dwarf spectroscopic classification system of \citet[][based on lower-resolution spectra, but extending to L9]{geballe_etal02} with the appropriately defined 2.20~$\micron$ methane index, the spectral type of HD~203030B would be L7.5$\pm$1.5 (methane index value of 1.13$\pm$0.07).  Both classifications are consistent with: (1) a visual comparison of the $K$-band spectrum of HD~203030B to the spectra of L6--T0 dwarfs (Fig.~\ref{fig_hd203030b_spectrum}), (2) the near-IR colors of HD~203030B ($J-H=1.28\pm0.56; H-K_S=0.64\pm0.16$; cf.\ Table~\ref{tab_photometry}), and (3) the empirical near-IR absolute magnitude vs.\ spectral type relations for L and T dwarfs from \citet{kirkpatrick_etal00} and \citet{vrba_etal04}.  Hence, the spectral type of HD~203030B is later than those of the coolest hydrogen-burning stars ($\sim$L4), and it is a brown dwarf.

Given that HD~203030B is younger than typical field brown dwarfs, a lower-than-dwarf surface gravity may make its spectral classification with respect to field dwarfs inadequate.  Our spectrum of HD~203030B does not allow us to assess its surface gravity empirically because it lacks features with well expressed gravity sensitivity, the only one at $K$ band being the \ion{Na}{1} 2.21~$\micron$ doublet which disappears after L2 in the L dwarfs \citep{mclean_etal03}.  Models predict that at the age of the primary (130--400~Myr; see \S\ref{sec_age} below) the surface gravity of HD~203030B is $\approx$0.5~dex lower than the gravities of $\sim$3~Gyr-aged field L dwarfs.  This may affect the strength of the observed CO and H$_2$O absorption, which are weakly sensitive to gravity.  Enhanced absorption by water vapor that produces peaked $H$-band continua is indeed observed in younger, $<$10--50~Myr old, L dwarfs with low surface gravities\citep{lucas_etal01, kirkpatrick_etal06}.  Our spectrum of HD~203030B does indeed exhibit a slightly peaked continuum shape near 2.1$\micron$---an effect has not been previously observed in L dwarfs.  However, the effect of low surface gravity on CO and H$_2$O absorption at $K$ band is likely too weak by the age of HD~203030B, as it is not distinctive even in the younger (1--50~Myr) early-L dwarf 2MASS~J01415823--4633574 \citep{kirkpatrick_etal06}.  Moreover, our estimate of the spectral type of HD~203030B does not depend on CO and H$_2$O absorption indices, but on the CH$_4$ index, which, in the lack of methane absorption, measures the slope of the continuum at 2.2~$\micron$---far from any CO and H$_2$O bandheads.  The peaked appearance of our $K$-band spectrum near 2.1$\micron$ could, on the other hand, can be traced to a possible chromatic effect in one of the telluric spectra at wavelengths $<$2.1$\micron$.  Therefore, we believe that our estimate of the spectral type of HD~203030B is not affected by its potentially lower surface gravity.

\section{PROPERTIES OF HD 203030B: AGE, LUMINOSITY, MASS, AND EFFECTIVE
TEMPERATURE}
\label{sec_properties}

Fundamental sub-stellar parameters are degenerate with age.  Ages, masses, effective temperatures, and radii are empirically unconstrained in isolated brown dwarfs.  Only sub-stellar bolometric luminosities can be estimated from measurements of trigonometric parallaxes.
The brown dwarf HD~203030B offers an opportunity to further resolve some of these degeneracies by the additional constraint on its age because of its association with HD~203030.  Below, we first discuss the factors that determine the age of the primary and hence, of the secondary, by assumption of coevality.  We then use our knowledge of the age and the heliocentric distance of HD~203030 to obtain model-dependent estimates of the mass and effective temperature of its sub-stellar companion.

\subsection{Age of the Primary \label{sec_age}}

HD~203030 is a G8~V star \citep{jaschek78} and its age can be constrained from a number of photometric and spectroscopic indicators established from G stars in well-studied open clusters of known ages.  A probabilistic estimate of the age of HD~203030 can also be made based on comparison of its galactic space motion to the space motions of known stellar populations and moving groups.  The relevant stellar age indicators and the inferred ages are discussed below.

\paragraph{Position on the Color-Magnitude Diagram.}

HD~203030 has an apparent $V$-band magnitude of 8.48~mag \citep{eggen64} and a {\sl Hipparcos} parallax of $24.48\pm1.05$~mas \citep{perryman_etal97}.  Thus, given an absolute $V$-band magnitude of 5.42~mag and a $B-V$ color of 0.75~mag \citep{eggen64}, HD~203030 sits on the main sequence, and is between 0.1~Gyr and 10~Gyr old.

\paragraph{Stellar Rotation.}

Stellar rotation periods increase with age as a result of angular momentum transfer outward from the core to the envelope, and then via loss to stellar winds or to a circumstellar disk \citep[e.g.,][]{kraft70, kawaler88, pinsonneault_etal89}.  \citet{koen_eyer02} find a photometric period of 4.1~days for HD~203030 in the {\sl Hipparcos} data.  This period lies in between the rotation periods of late-G stars in the $\alpha$~Per \citep[0.2--2.4~days;][]{prosser92} and in the Coma / Hyades clusters \citep*[6--12~days;][]{radick_etal90}, and is within the range of G-star rotation periods observed in the Pleiades (0.4--6~days; \citealt*{vanleeuwen_etal87}; \citealt{krishnamurthi_etal98}).   Assuming that the photometric periodicity of HD~203030 is due to star spots and stellar rotation, the age of HD~203030 is between that of $\alpha$~Per \citep[90~Myr;][]{stauffer_etal99} and that of the Hyades \citep[625~Myr;][]{perryman_etal98}, and possibly comparable to the age of the Pleiades \citep[125~Myr;][]{stauffer_etal98}.

\paragraph{Chromospheric \ion{Ca}{2} H \& K Emission.}

The strength of \ion{Ca}{2} emission at 3968\AA\ and 3933\AA\ in the cores of the H \& K chromospheric absorption bands is a well-known measure of chromospheric activity, and a proxy for the evolution of the stellar dynamo.  A widely adopted indicator is the \Rhk\ index, which measures the relative strength of \ion{Ca}{2} emission with respect to the stellar continuum \citep{noyes_etal84}.  The value of the \Rhk\  index for HD~203030 has been obtained from high-resolution optical spectra of the star by four separate groups, the individual measurements being: --4.20 \citep{strassmeier_etal00}\footnote{\citet{strassmeier_etal00} list $R_{HK}$, which we convert to \Rhk\ using the $B-V$ color of HD~203030 and an expression from \citet{hartmann_etal84}.}, --4.36 (D.~Soderblom, 2000, private communication), --4.37 \cite{wright_etal04}, and -4.47 (White, Gabor, \& Hillenbrand, submitted).  The mean value of the index from the four sources is $\log \Rhk=-4.35\pm0.06$ (standard error).  Using an empirical age-\Rhk\ relation from \citet{donahue93},
\begin{equation}
\log (t) = 10.725 - 1.334 R_5 + 0.4085 R_5^2 - 0.0522 R_5^3 \label{eqn_rhk},
\end{equation}
where $t$ is the stellar age in years and $R_5 = 10^5 R_{HK}^\prime$, we calculate that the age of HD~203030 is $t=180^{+260}_{-140}$~Myr.  

\paragraph{Coronal X-Ray Emission.}

Similarly to chromospheric activity, coronal x-ray luminosity is another proxy for the strength of the stellar dynamo that can be applied to determine stellar ages \citep{vilhu84, randich00}.  In a study of age diagnostics for young Sun-like (G and early K) stars, Hillenbrand et al. (in preparation) find, based on an analysis of open cluster data, that the age of Sun-like stars is related to their x-ray luminosity following
\begin{equation}
\log (t_{\rm Myr}) = -0.88 \log L_X + 33.88 \label{eqn_lx},
\end{equation}
where $t_{\rm Myr}$ is the stellar age in Myr and $L_X$ is the x-ray luminosity in erg~s$^{-1}$.  The x-ray luminosity of HD~203030 from the {\sl Rosat} All-Sky Survey Faint Source Catalog \citep{voges_etal00} is $\log (L_X)=29.01\pm0.16$, where we have converted count rates to flux using the conversion factor in \citet{fleming_etal95b}.  From this, we find that the age of HD~203030 is $220^{+140}_{-90}$~Myr, where the intrinsic scatter of relation~(\ref{eqn_lx}), 0.16~dex (Hillenbrand et al., submitted), has been added in quadrature to the error in $L_X$. 

\paragraph{\ion{Li}{1} Absorption.}

Lithium is destroyed in the convective envelopes of late-type stars, and so a high photospheric lithium abundance is a signature of stellar youth \citep{herbig65, bodenheimer65}.  \citet{strassmeier_etal00} measure a \ion{Li}{1}~$\lambda6708$~\AA\ equivalent width of 56~m\AA\ for HD~203030 in $R=25000$ spectra.  This is a factor of $\sim$2 larger than the measured \ion{Li}{1} equivalent widths in $R\sim34000$ spectra of stars of similar colors in the 625 Myr-old Hyades \citep{soderblom_etal90}, a factor of $\sim$3 smaller than the \ion{Li}{1} equivalent widths in $R\sim20000$ spectra of late-G stars in the 125~Myr-old Pleiades \citep{duncan_jones83}, and is within the range of \ion{Li}{1} equivalent widths in $R\sim48000$ spectra of analogous stars in the M~34 (NGC~1039) open cluster \citep[250~Myr;][]{jones_etal97}.  HD~203030 is therefore younger than 625~Myr, probably older than 125~Myr, and likely of order 250~Myr old.

\paragraph{$UVW$ Galactic Space Motion.}

Young stars inherit the kinematics of their birth environment for tens to hundreds of Myr before their space motion is disrupted by dynamical interactions with the galactic environment.  The galactic velocity vector of a star can therefore be used to obtain a probabilistic estimate of the star's age, when compared to the space motions of comoving groups of stars with known ages.  Based on its galactic space motion ($U=-23.4$~km~s$^{-1}$, $V=-15.9$~km~s$^{-1}$, $W=-12.2$~km~s$^{-1}$), \citet{montes_etal01a} classify HD~203030 as a probable kinematic member of the IC~2391 super-cluster, with an estimated age of 35--50~Myr \citep*{mermilliod81, barradoynavascues_etal04b}.

To summarize, based on its chromospheric and coronal activity, HD~203030 is $\sim$250~Myr old, with a possible age range of 130--400~Myr.  This age is fully consistent with its stellar rotation rate and photospheric lithium absorption, and is also supported by its main-sequence position in a color-magnitude diagram.  However, this age does not agree with the young kinematic age of HD~203030 quoted in \citet{montes_etal01a}.  Nevertheless, such discrepancies are not unusual in kinematically selected samples of young stars.  Indeed, we note that the $UVW$ velocity vector of HD~203030 inhabits an intermediate position between the loci of the 20--150~Myr Local Association, the 100--300~Myr Castor moving group, and the 600 Myr Hyades super-cluster \citep[Fig.~1 in][and references therein]{montes_etal01a}.  If HD~203030 were a member of the Local Association or the Castor moving group, instead of the IC~2391 moving group, its kinematic age would be consistent with the ages derived from the other criteria.

\subsection{Luminosity \label{sec_L}}

We determine the absolute $K_S$-band magnitude of HD~203030B (Table~\ref{tab_photometry}), by assuming that it is at the same heliocentric distance as the primary \citep[{\sl Hipparcos} parallax $24.48\pm1.05$~mas;][]{perryman_etal97}.  To convert the absolute magnitude to bolometric luminosity, we use a calibration of the $K$-band bolometric correction BC$_K$ vs.\ L--T spectral type from \citet{golimowski_etal04}.  We take into account that the calibration of \citet{golimowski_etal04} is for the spectral classification system of \citet{geballe_etal02}, on which the spectral type of HD~203030B is L7.5$\pm$1.5 (\S\ref{sec_sptype}).  We also note that the BC$_K$ vs.\ spectral type relation from \citeauthor{golimowski_etal04} is for the MKO near-IR filter set, while our data are on the CIT filter set.  However, the synthetic transformations of \citet{stephens_leggett04} show that the MKO $K$ and CIT $K_S$ filters produce nearly identical results with an offset of 0.00--0.02~mag for L6--L9 dwarfs.  This small offset has been added to our error budget.  Finally, we note that the bolometric corrections in \citeauthor{golimowski_etal04} are compiled from data for $>$1~Gyr-old, high-surface gravity field dwarfs, and may need to be corrected for the expected $\approx$0.5~dex lower surface gravity of the $\sim$250~Myr-old HD~203030B.  The sense of this correction is unknown empirically, as the body of data on young L dwarfs is extremely limited.  From the models of \citet{burrows_etal97, burrows_etal01} and \citet{chabrier_etal00} we infer that the magnitude of the correction over a 0.5~dex change in surface gravity is $\leq$0.05~mag, and that it may be either positive or negative.  Therefore, we treat it as an added error term.

Averaging the bolometric corrections for the six L6--L9 dwarfs in Table~6 of \citet{golimowski_etal04}, we find BC$_K=3.19\pm0.11$~mag for HD~203030B.  The bolometric magnitude of HD~203030B is then $\Mbol= M_{K_S} + {\rm BC}_K = 16.34\pm0.18$~mag and, assuming $M_{{\rm bol},\sun}=4.74$~mag for the Sun \citep{drilling_landolt00}, its luminosity is $\log(L/L_\sun)=-4.64\pm0.07$.  The bolometric luminosity of HD~203030B is thus comparable to those of single field L6--T5 dwarfs (mean $\log(L/L_\sun)=-4.58$ with a standard deviation of 0.15~dex; cf.\ Table~6 in \citeauthor{golimowski_etal04}).  That is, at the 130--400~Myr age of HD~203030B, transition L/T dwarfs are not more luminous than their $>$1 Gyr-aged field counterparts.  This is an unusual but significant result, given that at its $\approx$0.5~dex lower anticipated surface gravity the luminosity of HD~203030B is expected to be $\approx$0.5~dex higher than that of older L/T dwarfs in the field (provided that the effective temperature at the L/T transition is independent of age).  As we discuss in the following sections, this translates in an unusually low effective temperature for HD~203030B (\S\ref{sec_teff}), and indicates either a necessity to re-evaluate the effective temperatures and ages of L/T dwarfs in the field or a gravity dependence in the spectral type---effective temperature relation near the L/T transition (\S\ref{sec_discussion_teff}).

\subsection{Effective Temperature \label{sec_teff}}

The effective temperature of HD~203030B can be inferred from its age and bolometric luminosity by adopting an age-dependent theoretical estimate of its radius.  Using the age and luminosity values obtained above, the models of \citet{burrows_etal97} predict a mean radius of 1.06~\Rj\ (at the the mean age and luminosity of HD~203030B) with a range of 1.00--1.22~\Rj\ (corresponding to the full age range and 1$\sigma$ luminosity range of HD~20303B), and a mean effective temperature of 1206~K with a range of 1090--1280~K.  These values and their ranges are also consistent with those inferred from the models of \citet[][DUSTY]{chabrier_etal00} and \citet[][COND]{baraffe_etal03}.

Alternatively, the effective temperature of HD~203030B can be inferred from its spectral type, by comparison to the compilation of spectral type and effective temperature data for brown dwarfs with known parallaxes in \citet{vrba_etal04} and \citet{golimowski_etal04}.  This seemingly empirical approach is also model-dependent because, despite the use of measured bolometric luminosities, \citet{vrba_etal04} and \citet{golimowski_etal04} also adopt brown-dwarf radii based on theoretical evolutionary models (from \citeauthor{burrows_etal97}).  Nevertheless, because these authors use only ensemble mean values for the radii \citep[$0.90\pm0.15$~\Rj;][]{vrba_etal04} or the ages \citep[3~Gyr;][]{golimowski_etal04} of field brown dwarfs, their results do not hinge on the details of the theoretical models.  From 12 distinct L dwarfs listed in Table~7 of \citet{vrba_etal04} and Table~6 of \citet{golimowski_etal04}, we find that the effective temperatures of L dwarfs of spectral types L7 or later (i.e., consistent with that of HD~203030B) fall in the range 1330--1650~K, with a mean of $1442\pm103$~K \citep[where we have excluded the temperature estimate for SDSS~J042348.57--041403.5, now resolved into a L6/T2 binary;][]{burgasser_etal05}.  The individual estimates have errors of approximately 100--200~K, arising from a combination of uncertainties in the photometry, parallaxes, and adopted radii or ages of the brown dwarfs.  

We note that the two ranges obtained for the effective temperature of HD~203030B, 1090--1280~K, based on its luminosity and the assumption of a model-dependent radius, and 1330--1650~K, from its spectral type and a comparison with the semi-empirical calibrations of \citet{vrba_etal04} and \citet{golimowski_etal04}, are only marginally consistent with each other, with HD~203030B appearing on average 240~K cooler.  The lower inferred effective temperature of HD~203030B is a direct consequence of the assumption of a $\approx$15\% larger radius (because of its young age) and similar bolometric luminosity (\S\ref{sec_L}) compared to late-L dwarfs in the field.  

\subsection{Mass \label{sec_mass}}

\begin{figure}
\plotone{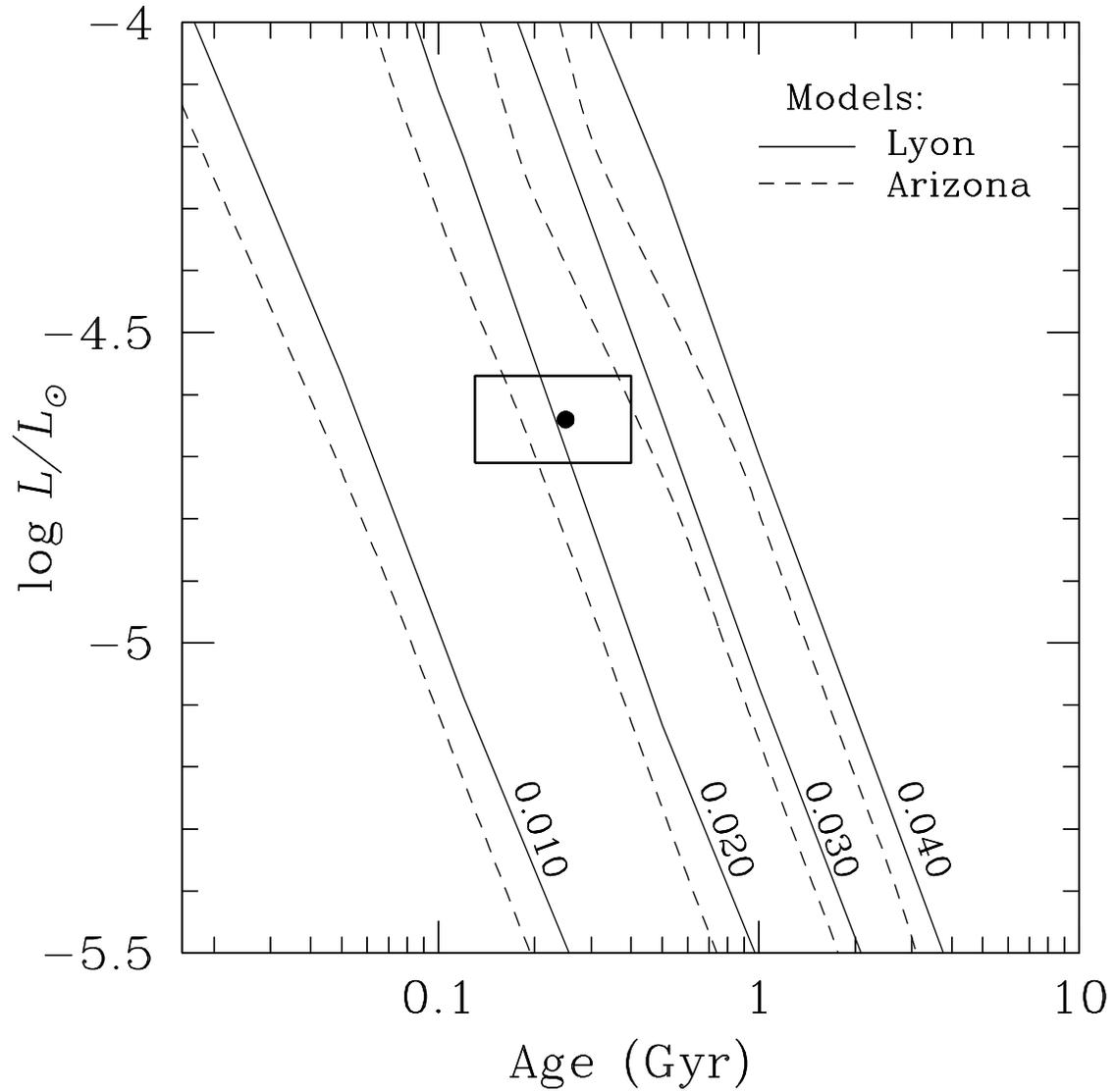}
\figcaption{A comparison of the luminosity and age of HD~203030B to predictions for 0.010--0.040~\Msun\ brown dwarfs from the evolutionary models of \citet[][solid lines]{baraffe_etal03} and \citet[][dashed lines]{burrows_etal97}.  The estimated mass of the sub-stellar companion is between 0.012~\Msun\ and 0.031~\Msun.
\label{fig_hd203030b_mass}}
\end{figure}

We estimate the mass of HD~203030B using the sub-stellar evolution models from the Lyon \citep{chabrier_etal00, baraffe_etal03} and Arizona \citep{burrows_etal97, burrows_etal01} groups.  Figure~\ref{fig_hd203030b_mass} shows the allowed locus of values (thick rectangle) around the adopted age and luminosity (solid dot) of HD~203030B overlaid on the evolutionary tracks.  We have only plotted the COND suite of models \citep{baraffe_etal03} from the Lyon group, as the evolution of sub-stellar luminosity predicted by the DUSTY \citep{chabrier_etal00} and COND models is nearly identical \citep{baraffe_etal03}.  Based on these predictions, the mass of HD~203030B is between 0.012~\Msun\ and 0.031~\Msun, with the adopted mean luminosity and age values corresponding to 0.023~\Msun.  If, instead, we estimate the mass of HD~203030B from the effective temperature ($1442\pm103$~K) corresponding to its spectral type from the calibrations of \citet{golimowski_etal04} and \citet{vrba_etal04}, then its mass would be $0.032^{+0.009}_{-0.012}$~\Msun.  However, the predicted bolometric luminosity of HD~203030B in this case would be a factor of 1.3--2.3 greater than the inferred one.

At a mass ratio of only $q=M_2/M_1\approx$0.02, the HD~203030~A/B system is among the lowest mass ratio binaries discovered through direct imaging to date, along with HR~7328~A/B \citep{lowrance_etal00}, AB~Pic~A/B \citep{chauvin_etal05b}, and GQ~Lup~A/B \citep{neuhauser_etal05}.  All of these very low mass ratio ($q\leq0.03$) binaries are young ($\leq$400~Myr) and have projected separations $>$100~AU, with estimated orbital periods near the $\sim$$10^5$-day peak of the G-star binary period distribution of \citet{duquennoy_mayor91}.  
Wide very low mass ratio systems may thus not be highly unusual at young ages and will be important for understanding the architecture of planetary systems in conjunction with their closer-in counterparts found from radial-velocity surveys.

\section{THE EFFECTIVE TEMPERATURE AT THE L/T TRANSITION}
\label{sec_lt_teff}

The uncharacteristically low effective temperature of the young HD~203030B sets it apart from older L7--L8 field dwarfs with known parallaxes.  Because the age of HD~203030B is known, unlike the ages of field brown dwarfs, this result has the potential of providing an important constraint for understanding the physics of the L/T transition.  However, prior to delving into the implications, we critically consider all factors underlying our estimate of the effective temperature of HD~203030B.

\subsection{Reliability of the Effective Temperature Estimate \label{sec_discussion_reliability}}

Our estimate of the effective temperature of HD~203030B hinges on the following assumptions and factors:
\begin{enumerate}
\item physical association and coevality of HD~203030A and B;
\item accuracy of the photometry of HD~203030A, B, and of the reference HD~13531A, or variability in any of the above;
\item accuracy of the spectral classification;
\item age of HD~203030A;
\item heliocentric distance of HD~203030A;
\item unknown multiplicity of HD~203030B;
\item bolometric correction for HD~203030B;
\item accuracy of the theoretical evolutionary models.
\end{enumerate}

We critically address each of these issues below.
\begin{enumerate}
\item If HD~203030B was older than the primary, the theoretical estimate of its radius could be sufficiently small ($\lesssim$0.9~\Rj) to drive its mean effective temperature (corresponding to its mean age and luminosity; \S\S\ref{sec_age}, \ref{sec_L}) above 1330~K and to make it consistent with the effective temperatures of L/T dwarfs in the field.  However, the physical association of HD~203030B with its primary was already demonstrated beyond reasonable doubt in \S\ref{sec_astrometry}.  While the possibility remains that HD~203030B is an old field brown dwarf captured by a younger HD~203030A, it is highly unlikely.

\item An overestimate of the apparent brightness of HD~203030B by $\geq$0.3~mag would be sufficient to resolve the effective temperature discrepancy.  This remains a possibility, especially given that precise ($\leq$0.1~mag) AO photometry is notoriously difficult to obtain because of variations in the quality of the AO PSF on time-scales of seconds.  Our observations on 28 August, 2002, benefitted from photometric conditions and relatively stable AO performance, with $K_S$-band Strehl ratios ranging between 40\% and 60\%.  Furthermore, our choice of wide ($\approx$8~FWHM) photometric apertures alleviates the effect of variations in the total enclosed power that might be caused by changes in the AO correction.  Both of these factors contributed to the high degree of self-consistency of our $K_S$-band AO photometry: $\leq$0.05~mag among individual exposures (\S\ref{sec_phot}).  We therefore believe that a systematic error in our photometry is not the culprit for the inferred discrepancy in the effective temperature.

Even though the photometry may be accurate, any of HD~203030A, HD~203030B, or HD~13531A (the photometric reference), could be variable by $\geq$0.3~mag at $K_S$, and thus possibly account for the fact that HD~203030B appears cooler than expected (because of being under-luminous).  Indeed, HD~203030A and HD~13531A are both young ($\leq$0.5~Gyr) stars and therefore expected to be variable to some extent.  Furthermore, variability amplitudes as high as 0.5~mag have been observed in L/T transition dwarfs on time-scales of 0.3--3~hours \citep*{enoch_etal03}.  However, as already pointed out (\S\ref{sec_phot}), the {\sl Hipparcos} photometry of HD~13531 shows no variability, hence HD~13531 appears to be an adequate photometric standard.  In addition, our differential photometry between HD~203030B and ``object~1'' from the individual 60~sec $K_S$-band coronagraphic exposures spanning 1.3~hours on Aug 28 2002 does not exhibit deviations larger than 0.1~mag from the mean (r.m.s.\ scatter~= 0.06~mag).  Barring the highly unlikely scenario in which HD~203030B and ``object~1'' follow photometric variations of similar magnitudes, periods, and phases, we infer that HD~203030B is not strongly variable, and therefore its low effective temperature cannot be explained by a period of low intrinsic luminosity.

\item Spectral classification of L dwarfs in the near-IR is still not as well established as in the visible.  Optically-typed L dwarfs do not follow a unique and monotonic spectral sequence in the near-IR, indicating that the two wavelength regions probe different photospheric physics \citep{kirkpatrick05}.  It is therefore possible that the spectral type of HD~203030B inferred from its $K$-band spectrum (Fig.~\ref{fig_hd203030b_spectrum}) does not correspond to the same spectral type in the visible.  However, there is very little room for error in the present spectral classification.  On one hand, the spectral type HD~203030B is not later than L8 \citep[on the L dwarf classification scheme of][]{kirkpatrick_etal99}, i.e., it is not a T dwarf, because its $K$-band spectrum lacks methane absorption at 2.20~$\micron$ (\S\ref{sec_sptype}).  On the other hand, if the spectral type of HD~203030B was earlier than L7, its low effective temperature would be even more discrepant with those of similarly-type field L dwarfs.  We are therefore confident that our $K$-band spectral classification of HD~203030B is accurate to the claimed precision.
 
\item An older age for HD~203030 would imply a longer contraction time for the sub-stellar companion and hence, a smaller radius.  Using the models of \citeauthor{burrows_etal97}, we find that, given the measured bolometric luminosity, an age $>$1~Gyr would be required at to make the brown dwarf radius sufficiently small ($\lesssim$0.9~\Rj) to raise its effective temperature above 1330~K.  However, as discussed in \S\ref{sec_age}, all age-dating criteria indicate that HD~203030 is younger than the Hyades (625~Myr).

\item If HD~203030 was at a larger heliocentric distance, HD~203030B would be intrinsically more luminous, and hence, hotter.  A 13\% error in the {\sl Hipparcos} parallax could account for the observed discrepancy in effective temperatures.  We note that this would be of the same relative magnitude and sense as the error in the {\sl Hipparcos} distance to the Pleiades \citep{pinsonneault_etal98}.  It is possible that such systematic zonal errors of similar magnitude exist also in other parts of the sky \citep[e.g., the Coma cluster; ][]{pinsonneault_etal98}.  However, the inferred {\sl Hipparcos} systematic error leading to the Pleiades distance problem is of order 1~mas.  This would result in only a $\approx$4\% error in the parallax (24.48~mas) of HD~203030: too small to explain the effective temperature discrepancy.

\item The existence of an unresolved binary companion to HD~203030B, a hypothetical HD~203030``C'' at $\lesssim$4~AU from the secondary, would change the inferred bolometric luminosity of the secondary.  Indeed, previously unknown binarity at the L/T transition has been recently invoked as a possibility in explaining the observed brightening in $J$ absolute magnitude from late-L to early-T spectral types \citep{burgasser_etal05, liu_etal05}.  However, a companion to HD~203030B would have the opposite effect: it would decrease the bolometric luminosities of the individual components (HD~203030B and ``C''), without much affecting their spectral types, and would thus exacerbate the problem.

\item In estimating the absolute $K_S$-band magnitude of HD~203030B we used bolometric corrections from \citet{golimowski_etal04}.  The accuracy of the bolometric corrections is limited by the presently small amount of data on L and T dwarfs and by the even smaller body of knowledge of the non-Raleigh-Jeans IR spectral energy distributions of brown dwarfs at $>$3~$\micron$.  Only recently has {\sl Spitzer}/IRS spectroscopy over 5.5--38$\micron$ confirmed that the bolometric corrections of \citeauthor{golimowski_etal04} are accurate to within 0.10~mag \cite{cushing_etal06}, albeit for a small number of L and T dwarfs.  Regardless, systematic uncertainties in the $K$-band bolometric corrections of L and T dwarfs do not affect our comparison of effective temperatures in relative terms, because both studies that calibrate L and T spectral types vs.\ effective temperature \citep{vrba_etal04, golimowski_etal04} use the same scale of bolometric corrections as adopted by us.  

\item Finally, our calculation of the effective temperature for HD~203030B hinges on the accuracy of the sub-stellar evolutionary models of \citet{burrows_etal97, burrows_etal01}.  Without further observational constraints (e.g., dynamical mass, surface gravity) it is not possible to empirically test the accuracy of these theoretical predictions.  However, the radii of the brown dwarfs in the studies of \citet{vrba_etal04} and \citet{golimowski_etal04} are also estimated based on these models.  Our approach in estimating the effective temperature of HD~203030B from its bolometric luminosity and age is therefore fully consistent with the methods employed in these two studies.
\end{enumerate}
Having considered and rejected all possible factors that may have lead us to underestimate the effective temperature of HD~203030B ($1206^{+74}_{-116}$~K) in comparison to the temperature expected from its L7.5$\pm$0.5 spectral type ($1442\pm103$~K), we conclude that HD~203030B is indeed cooler than late-L dwarfs in the field.

\subsection{Origin of the Temperature Discrepancy \label{sec_discussion_teff}}

We pointed out in \S\ref{sec_discussion_reliability} that the discrepantly low effective temperature of HD~203030B arises mostly from the larger sub-stellar radius adopted for HD~203030B, because of its younger age but similar luminosity compared to older brown dwarfs of the same spectral type.   Having established the robustness of our age and luminosity determinations of HD~203030B (\S\ref{sec_discussion_reliability}), we conclude that either: (1) the effective temperatures of brown dwarfs across the L/T transition in \citet{golimowski_etal04} and \citet{vrba_etal04} have been over-estimated on average by 240~K because of unresolved binarity, (2) the ages of L/T transition dwarfs in the field have been over-estimated, or equivalently, their radii under-estimated (and hence, their effective temperatures over-predicted) by theoretical models, or (3) the range of effective temperatures over which the L/T transition occurs extends to cooler temperatures than found in the empirical studies of \citet{golimowski_etal04} and \citet{vrba_etal04}, e.g., due to variations in surface gravity.  We address each of these possibilities in turn below.

The effective temperatures of field brown dwarfs at the L/T transition may have been over-estimated if the majority of these brown dwarfs are unresolved close binaries, i.e., if they appear over-luminous.  Indeed, \citet{burgasser_etal05} find that $\sim$50\% of late-L to early-T dwarfs turn out to be binaries when imaged at high angular resolution (e.g., with AO, or with {\sl HST}), including one of the L/T brown dwarfs considered in the studies of \citet{golimowski_etal04} and \citet{vrba_etal04}, SDSS~J042348.57--041403.5.  We test if unresolved binarity is a factor by focusing only on these L/T transition dwarfs that have been imaged at high angular resolution.  Seven of the L7--L8 dwarfs in \citeauthor{vrba_etal04} have a priori known multiplicity from {\sl HST}/WFPC2 \citep{bouy_etal03, gizis_etal03}, i.e., they have either been resolved into binaries, or have remained unresolved.  The mean effective temperature of these is $1459\pm102$~K.  It is similar to, if not somewhat higher than, the mean effective temperature, $1403\pm117$~K (both sets of errors denote standard deviations), of the remaining four L7--L8 dwarfs that have not been observed at high angular resolution.   Therefore, it is unlikely that unresolved binarity has lead to a significant over-estimate of the effective temperatures across the L/T transition.  

The second hypothesis (overestimated ages leading to underestimated radii) is not directly testable with the existing observational data, as the radii of L and T dwarfs have not been measured empirically.  Nevertheless, following our discussion in \S\ref{sec_teff}, we construct an indirect test, by comparing the effective temperatures of all known L7--L8 dwarf companions to stars (with known ages and distances) to the effective temperatures of field late-L dwarfs with known trigonometric parallaxes from the studies of \citeauthor{golimowski_etal04} and \citeauthor{vrba_etal04}  Given a self-consistent application of the theoretical models \citep{burrows_etal97, burrows_etal01}, the two sets of late-L dwarfs, companions vs.\ free-floating, should have identical effective temperatures.  The inferred model-dependent effective temperatures of all L7--L8 companions to stars (GJ~584C, \citealt{kirkpatrick_etal01}; GJ~337C, \citealt{wilson_etal01, burgasser_etal05b}; HD~203030B, this paper) are all $\leq$1345~K, which is $\geq$100~K below the mean for field L7--L8 dwarfs with known trigonometric parallaxes \citep[\S\ref{sec_teff};][]{golimowski_etal04, vrba_etal04}.  While the corresponding deviation in each case is only 1--2$\sigma$, the statistical significance of the overall effect is $>$99.5\% (i.e., $\gtrsim$3$\sigma$).  If the effect is real, it indicates that the predicted radii of field late-L dwarfs are 10--15\% too small (making their effective temperatures too high) compared to those of the late-L companions.  Equivalently, this translates into a factor of 1.5--2 over-estimate of the adopted ages for late L dwarfs in the field, according to the models.  Indeed, the mean ages of the three L7--L8 companions (ranging between 0.3 and 2.0~Gyr) are a factor of $\gtrsim$1.5 smaller than the 2.9 Gyr mean statistical age of late-L dwarfs in the field (\citealt{allen_etal05}; also based on models from \citealt{burrows_etal01}).  While the youngest of the late-L companions, HD~203030B, was the product of a targeted search for brown dwarfs around $\leq$0.5~Gyr old stars \citep{metchev_hillenbrand04, metchev06}, and so should be excluded from this age comparison, GJ~584C and GJ~337C were not.  Thus, barring a systematic under-estimate of the mean ages of these two stellar primaries by a factor of $\sim$1.5 \citep[not at all unlikely, given that their maximum ages are 2.5 and 3.4~Gyr, respectively;][]{kirkpatrick_etal01, wilson_etal01}, it is possible that the ages of field late-L dwarfs have been overestimated by a factor of $\sim$1.5.  A similar age overestimate is inferred when using the models of \citet{chabrier_etal00} and \citet{baraffe_etal03}.  That is, the result does not hinge on peculiarities in either set of models.  Thus, sub-stellar cooling rates may have been under-predicted, or effective temperatures in the initial conditions may have been over-estimated, in both the Arizona and the Lyon models.   However, given the strong reliance of this result on the ages of two isolated stars, which are not determined as accurately as the ages of stars in stellar associations, we cannot claim to have discovered any inaccuracy in the theoretical models with certainty.  The hypothesis needs to be re-visited once accurate bolometric luminosities are determined for a larger number of sub-stellar objects with well-known ages, e.g., in open clusters.

The last hypothesis (that the L/T transition extends to cooler temperatures) is the default one, as it reconciles previous determinations of the effective temperature at the L/T transition with the present data, without seeking constraints on the underlying physics.  A physical interpretation may nonetheless focus on the role of metallicity and surface gravity in the L/T transition \citep*[e.g.,][]{burrows_etal06}.  The metallicity of HD~203030B may be assumed to be the same as that of the primary, for which \citet{nordstrom_etal04} measure ${\rm [Fe/H]}=-0.01$, i.e., solar, and likely not very different from that of most field L dwarfs.  However, as already discussed (\S\ref{sec_sptype}), the gravity of the $\sim$250~Myr-old HD~203030B is expected to be $\approx$0.5~dex lower than the gravities of $>$1~Gyr-old field brown dwarfs, as corresponding to its 10--15\% larger adopted radius.  While we determined that gravity has not affected our estimates of the effective temperature and luminosity of HD~203030B in a systematic way, its effect on photospheric chemistry in brown dwarfs is unknown.  We note that, among stars, late-type giants have 200--600~K lower effective temperatures than dwarfs of the same spectral type \citep[c.f.,][]{gray92a, dyck_etal96}.  Hence, a similar trend among sub-stellar objects is not at all unlikely.  A dependence on surface gravity in the effective temperature at which the L dwarf sequence merges onto the T dwarf sequence is indeed predicted by the ``unified cloudy model'' of \citet{tsuji_nakajima03}.  These authors propose such a dependence to explain the observed brightening of early-T dwarfs at $J$ band \citep{dahn_etal02, tinney_etal03, vrba_etal04}, arguing that the brighter early-T dwarfs have lower surface gravities because they are younger.  We note that \citeauthor{tsuji_nakajima03}'s theory does not agree with the fact that the young HD~203030B is not brighter than its older late-L counterparts in the field (\S\ref{sec_L}).  Furthermore, \citet{liu_etal06} and \citet{burgasser_etal06} have recently demonstrated that the $J$-band brightening across the L/T transition is not an age-spread effect, but a real feature of brown dwarf evolution, as might be explained by disruption of condensate clouds at a fixed effective temperature \citep{ackerman_marley01, burgasser_etal02b}.  Still, with HD~203030B we have now shown that the effective temperatures of different L/T transition objects may not be identical.  It therefore remains an open question whether surface gravity may have a role in setting the effective temperature at the L/T transition. Planned optical and $J$-band spectroscopy of HD~203030B, combined with theoretical modeling of gravity-sensitive alkali, metal-hydride, and VO features will attempt to address this issue.  Unfortunately, the magnitude of the effect that we will be trying to measure is of order of the accuracy with which of modern models of sub-stellar photospheres can match gravity-sensitive features in brown-dwarf spectra.  The role of gravity on the effective temperatures of L dwarfs at a fixed spectral type may thus be better addressed with younger (albeit earlier-type) objects, such as the recently discovered 2MASSW~J1207334--393254B \citep{chauvin_etal05a} and 2MASS~J01415823--4633574 \citep{kirkpatrick_etal06}.

\section{CONCLUSION}

We have discovered a proper motion sub-stellar companion to the 130--400~Myr-old Sun-like star HD~203030.  The companion, HD~203030B, is 11$\farcs$9 away from the primary, at a projected orbital separation of 487~AU.  Assuming coevality with the primary star, sub-stellar evolution models place the companion mass at $0.023^{+0.008}_{-0.011}\Msun$.  From $K$-band AO spectroscopy, we determine a spectral type of L7.5$\pm$0.5 for HD~203030B.  The brown dwarf is thus near the spectroscopic L/T transition, characterized by diminishing amounts of dust in sub-stellar photospheres.  Because of its association with a main-sequence star of a known age and heliocentric distance, HD~203030B offers a rare chance to probe the L/T transition in a setting in which most of the degeneracies characteristic of sub-stellar models are resolved.

We find that the bolometric luminosity of HD~203030B is comparable to that of $>$1~Gyr-old field dwarfs of similar spectral types, despite the $\approx$0.5~dex lower gravity anticipated at the young age of HD~203030B.  As a result, theoretical models of sub-stellar evolution predict a $\approx$240~K cooler effective temperature for HD~203030B compared to the effective temperatures of its older counterparts in the field.  We consider three hypotheses for the discrepancy: (1) that the bolometric luminosities and effective temperatures of late-L dwarfs in the field have been over-estimated because of unresolved binarity; (2) that the effective temperatures of late-L dwarfs in the field may have been over-estimated (equivalently, their radii under-estimated) because theoretical models may have over-predicted their ages; and (3) that the spectral type---effective temperature for late-L dwarfs is not age- and gravity-independent.   Based on multiplicity studies of field brown dwarfs from high-angular resolution observations with {\sl HST} and AO, we do not find a significant discrepancy between the effective temperatures (or luminosities) of brown dwarfs with known and unknown multiplicities, and rule out the first possibility.  We test the second hypothesis by comparing the effective temperatures of late-L secondaries of main-sequence stars (with known ages and distances) to late-L dwarfs in the field.  We find evidence that the late-L companions are $\geq$100~K cooler and $\sim$1.5 times younger than their isolated counterparts.  However, because ages of individual stars cannot be claimed with such accuracy, the significance of this result is only marginal.  The last hypothesis is a compromise between the discrepant effective temperatures of HD~203030B and late-L dwarfs in the field, and draws a parallel with the dependence of spectral type and effective temperature on surface gravity in stars \citep{gray92a}.  Both remaining hypotheses await testing from larger samples of brown dwarfs with known distances and ages, either as members of known stellar associations or as companions to stars.

\begin{acknowledgments}
The authors acknowledge many insightful discussions with Travis Barman, J.~Davy Kirkpatrick, and Ian McLean during the preparation of the manuscript.  We are grateful to our support astronomers and engineers Randy Campbell, Rick Burress, and Jeff Hickey for their guidance in using Keck and Palomar AO, to Keith Matthews and Dave Thompson for advice with NIRC2 spectroscopy, and to our telescope operators at the Palomar and Keck~II telescopes, Jean Mueller, Karl Dunscombe, and Cynthia Wilburn.  We also wish to extend special thanks to those of Hawaiian ancestry, on whose sacred mountain of Mauna Kea we are privileged to be guests.  Without their generous hospitality, the spectroscopic observations presented herein would not have been possible.  S.~A.~M. acknowledges support from a {\sl Spitzer Space Telescope} Postdoctoral Fellowship.   This work was supported by NASA grant NNG05GJ37G and NASA/JPL contract 1224566.
\end{acknowledgments}
 

\clearpage

\begin{deluxetable}{ccccc}
\tablewidth{0pt}
\tablecaption{Observations of HD 203030 \label{tab_observations}}
\tablehead{\colhead{Epoch (UT)} & \colhead{Photometric?} & \colhead{Observing Mode} & \colhead{Telescope}}
\startdata
2002 Aug 28 & yes & $J H K_S$ imaging & Palomar \\
2003 Jul 16 & no & $K_S$ imaging & Palomar \\
2004 Jun 26 & no & $K_S$ imaging & Palomar \\
2005 Jul 14 & yes &$J K_S$ imaging, $K$ spectroscopy & Keck \\
\enddata
\end{deluxetable}

\begin{deluxetable}{ccccc}
\tablewidth{0pt}
\tablecaption{Near-IR Photometry and  of HD~203030B \label{tab_photometry}}
\tablehead{\colhead{$J$} & \colhead{$H$} & \colhead{$K_S$} & \colhead{$\Delta K_S$}
	& \colhead{$M_{K_S}$}}
\startdata
$18.13\pm0.55$ & $16.85\pm0.12$ & $16.21\pm0.10$ & $9.56\pm0.10$ & $13.15\pm0.14$ 
\tablenotemark{\dag}
\enddata
\tablenotetext{\dag}{Assuming the {\sl Hipparcos} parallax of $24.48\pm1.05$~mas for HD~203030A \citep{perryman_etal97}.}
\end{deluxetable}

\clearpage

\epsscale{0.6}

\epsscale{1.0}

\end{document}